\documentclass[paper]{agujournal2019}
\usepackage{url}
\usepackage{lineno}
\usepackage{color}
\usepackage{amsmath}
\draftfalse

\newcommand{\ssr}{   {Space Sci. Rev. }}

\newcommand{\jgr}{   {J. Geophys. Res.}}
\newcommand{\grl}{   {Geophys. Res. Lett.}}
\newcommand{\apj}{   {Astrophys. J.}}
\newcommand{\apjl}{   {Astrophys. J. Lett.}}

\graphicspath{{./}{figures/}}

\journalname{JGR: Space Physics}

\begin{document}


\title{Whistler-mode waves in near-equatorial THEMIS measurements: reconstruction of magnetic field spectra from electric field and plasma measurements.}

\authors{Declan Frawley\affil{1}, Dmitri L. Vainchtein\affil{1}, Anton V. Artemyev\affil{2,3}, Vassilis Angelopoulos\affil{3}} 
\affiliation{1}{Nyheim Plasma Institute, Drexel University, Camden, NJ, USA}
\affiliation{2}{Department of Physics, University of Texas at Arlington, Arlington, TX, USA}
\affiliation{3}{Earth, Planetary, and Space Sciences, University of California, Los Angeles, Los Angeles, CA, USA}

\correspondingauthor{Dmitri Vainchtein}{dlv36@drexel.edu}

\begin{keypoints}
\item We propose a method of reconstruction of whistler-mode spectral density measured by THEMIS E and D after 2016 
\item Comparison with THEMIS A measurements shows the range of uncertainties of restored wave spectral density
\item We used the Z-Score method for isolation of whistler waves from the low-intensity noise.
\end{keypoints}

\begin{abstract}
Electromagnetic whistler-mode waves are a natural emission in the outer radiation belt and the Earth’s magnetotail. The resonant interaction of these waves and energetic electrons are responsible for electron acceleration and losses, thus coupling the magnetosphere and ionosphere. Near-equatorial spacecraft use search-coil magnetometers for whistler-mode wave measurements, and one of the largest (covering the longest period of time) dataset of such waves has been collected by the THEMIS mission operating in the near-Earth magnetosphere within 2008-2025. However, after 2017, the search-coil magnetometers on two THEMIS spacecraft, THEMIS E and D, experienced problems with their signal along the spacecraft spin axis and were only able to detect the spin plane components of the wave vector. This significantly reduces our ability to detect the total wave amplitude wave magnitudes and limits our ability to incorporate the THEMIS E, D datasets into investigation of whistler-mode waves. In this technical report, we propose and validate a technique for reconstruction of magnetic field spectral density for Fast Fourier transform data product collected during Fast-Survey mode hereafter referred to as the {\it fff} dataset  collected by THEMIS E and D. We use measurements of the electric field instrument and cold plasma dispersion relation to evaluate the whistler-mode magnetic field spectral density. Verification of this technique by comparison with THEMIS A measurements (which retained their 3D measurement capability intact) confirms that restored magnetic field spectral density is within a factor of $\times 1.5$ of the actually measured magnitudes.
\end{abstract}

\section{Introduction}  
Time History of Events and Macroscale Interactions during Substorms (THEMIS) mission started operating in the equatorial Earth's magnetosphere in 2007 \cite{Angelopoulos08:ssr}. Originally this mission consisted of five spacecraft (A-E). In 2009, two of THEMIS spacecraft (C and B) were moved to a lunar orbit, where they now form the ARTEMIS mission \cite{Angelopoulos11:ARTEMIS}, whereas THEMIS A, D, and E still operate in the near-Earth magnetosphere and in the solar wind (with the apogee about $12R_E$). The THEMIS spacecrafts are equipped with identical instruments needed for plasma and electromagnetic investigation. Specifically, there are search-coil magnetometers \cite{LeContel08} and electric field instruments \cite{Bonnell08} that provide measurements of high-frequency electromagnetic fields (up to $8$kHz). These measurements are processed onboard into Fourier spectra with high frequency resolution and $1$s cadence \cite<the so-called {\it fff} dataset; see>{Cully08:ssr}. Multi-spacecraft opportunities and an unprecedentedly long period of THEMIS operation within different magnetosphere regions (magnetotail, radiation belts, magnetopause, and bow shock) make the {\it fff} dataset very useful for investigation of statistical properties of electromagnetic waves \cite<e.g.,>{Agapitov18:correlations,Gao22:Luphi_whistlers}. These opportunities are further enhanced via THEMIS multiple conjugations with low-altitude missions \cite<e.g.,>{Zhang22:natcom,Tsai25:review} and ground-based stations \cite<e.g.,>{Artemyev21:jgr:ducts}.  

The frequency range of the {\it fff} dataset covers multiple wave-modes in different magnetosphere regions: whistler-mode chorus, hiss, and magnetosonic waves in the inner magnetosphere \cite{Chen12:fff,Ma13,Li15:fff}; electron cyclotron harmonics, whistler-mode waves, and kinetic Alfven waves in the magnetotail \cite{Zhang14:ECH,Zhang18:whistlers&injections,Chaston12}; whistler-mode waves, electron acoustic waves, and magnetosonic waves in the bow shock and solar wind \cite{Artemyev22:jgr:bowshock,Cattell20:shocks}. One of the most important and widespread modes covered by the {\it fff} dataset are whistler-mode waves propagating within the $[1/40,1]f_{ce}$ frequency range (where $f_{ce}$ is the electron gyrofrequency). Whistler-mode waves are generated by anisotropic electrons, often occupy a narrow frequency range associated with energies of specific (local) electron population \cite<see details in>[and references therein]{Frantsuzov22:pop,Roytershteyn24}, and are well detected in the {\it fff} spectra. Although the {\it fff} dataset does not contain information about whistler-mode propagation, \cite<some estimates can be done using spectra of different magnetic field components; see>{Tong19:ApJ}, the main parameter of wave models, the wave spectrum and intensity, can be derived.  

In 2017-2019, technical issues with the search-coil magnetometers on THEMIS E and D started revealing themselves. Instead of three magnetic field components used for constructing the {\it fff} spectra, only one (along the spacecraft spin; approximately along the $z$-gsm axis) shows a reasonable signal, whereas magnitudes of two other components are artificially smaller. This significantly affected the ability to determine the wave propagation direction \cite<see discussion in>{Zhang22:natcom} and require using electric field measurements to estimate wave intensity \cite<see discussion in>{Shen&Li23:elfin_ulf}. Therefore, using THEMIS E and D {\it fff} datasets after $\sim$2017 requires additional scaling of magnetic field wave intensity. In this technical report, we test a technique for scaling of field-aligned whistler-mode waves, the most intense and important wave mode for electron scattering and acceleration in the plasma sheet and the outer radiation belts. 

The idea of scaling grew from the experience with the CRESS satellite where only electric field measurements were available for the whistler-mode frequency range \cite<e.g.,>{Ni11}. A similar technique has been applied previously to specific datasets where only electric field measurements were available for whistler-mode waves \cite<e.g.,>{Agapitov14:jgr:acceleration,Ma17:vlf}. We used the original (with an underestimated intensity) {\it fff} dataset from the THEMIS E and D search-coil magnetometers to isolate the frequency range of whistler-mode waves, the electric {\it fff} dataset for the electric field spectra, plasma density measurements from spacecraft potential measurements \cite{Nishimura13:density}, electron gyrofrequency from flux-gate magnetometer measurements \cite{Auster08:THEMIS}, and cold plasma dispersion relation for whistler-mode waves \cite{bookStix62} to reconstruct the magnetic field intensity for different frequency channels for the {\it fff} dataset.  

The technical report has the following structure: In Section \ref{sec:data} we present two examples of the reconstruction technique for THEMIS E and THEMIS A {\it fff} datasets (we use THEMIS A measurements to assess the accuracy of reconstruction of magnetic field intensity), In Section \ref{sec:statistics} we discuss statistical properties of the reconstruction technique, and in Section \ref{sec:conclusions} we discuss the applicability range of this technique. We also include Appendix describing the Z-score method to filter the {\it fff} dataset and isolate intense bursts associated with whistler-mode waves.
 
\section{The reconstruction technique}\label{sec:data}
\begin{figure*}
\centering
\includegraphics[width=1\textwidth]{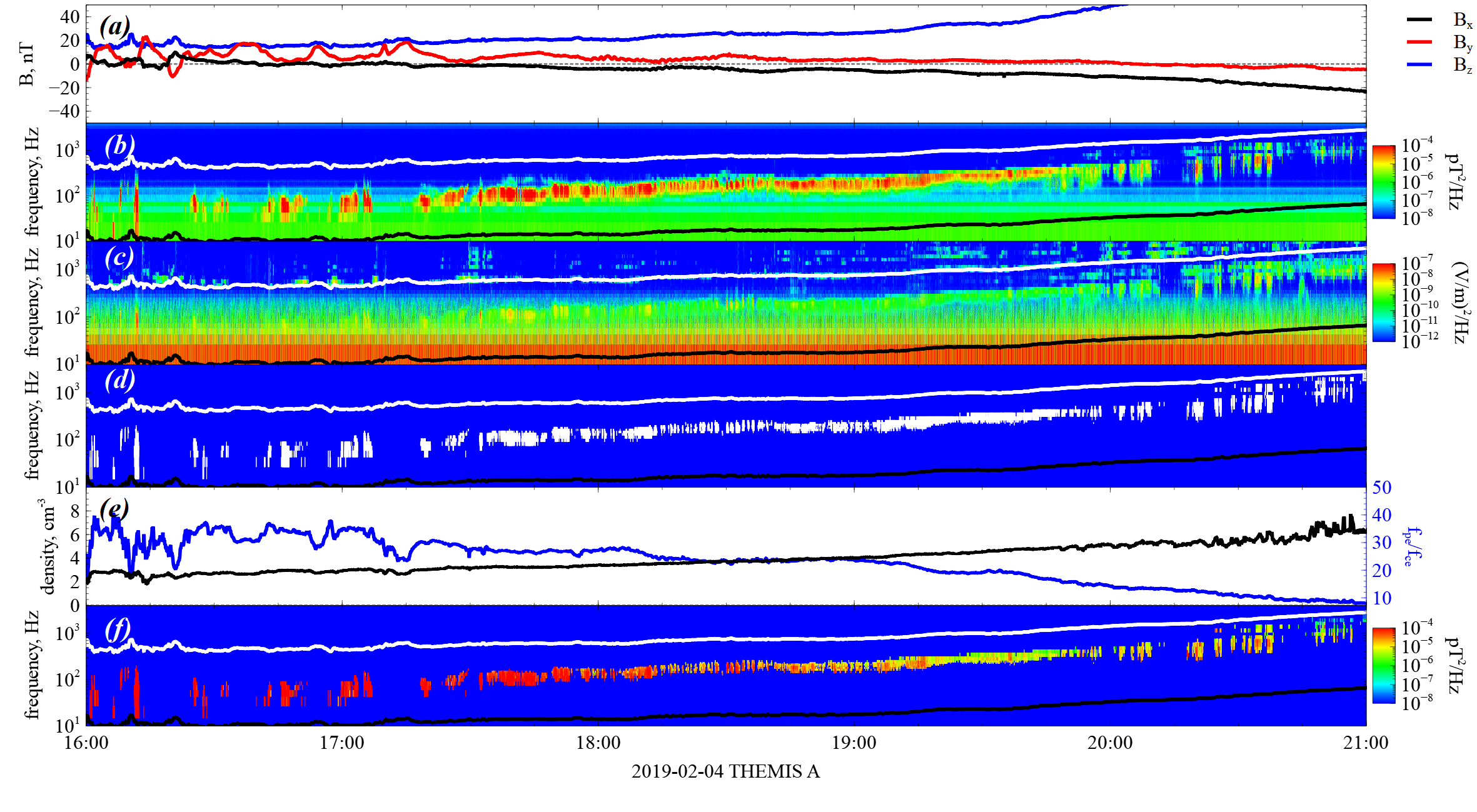}
\caption{Example of a {\it good} event from THEMIS A measurements with (a) magnetic filed (GSM coordinates), magnetic field spectra density from {\it fff} (b), electric field spectra density from {\it fff} (c), frequency domains of whistler-mode waves determined from the magnetic field spectra (d) plasma density and $f_{pe}/f_{ce}$ ratio (e), reconstructed magnetic field spectra density (f). White and blue regions in panel (d) can be considered as having values $1$ (waves) and $0$ (noise); then ``multiplication'' of (d) and (c) panels gives the cleaned electric field spectra associated with whistler-mode waves and can be further recalculated into magnetic field spectra shown in panel (f). } \label{fig1}
\end{figure*}

To demonstrate the reconstruction technique, let us consider a test dataset collected by THEMIS A with a well-functioning search-coil magnetometer. Figure \ref{fig1} shows an event (5 hours) from this dataset. These are equatorial measurements ($B_z$ is much larger than $B_x$, $B_y$; see panel (a)) of intense whistler-mode waves. Panel (b) shows the {\it fff} spectrum obtained from the search-coil magnetometer. Using magnetic field measurements, we can distinguish the intense waves within the whistler-mode frequency range, $[1/40,1]f_{ce}$, (see details in Appendix). Then, we clean the electric field {\it fff} dataset, keeping only waves associated with whistler-mode waves detected in the magnetic field {\it fff} dataset: compare frequency domains covered by waves in the original electric field {\it fff} spectrum in panel (c) and cleaned spectrum in panel (d). The electric field spectrum within the frequency domains from panel (d) is recalculated into a magnetic field spectrum (a reconstructed magnetic field spectrum) using the whistler-mode dispersion relation for cold plasma \cite{bookStix62,Ni11}. Such recalculation is performed for each frequency channel and involves local $f_{ce}$ and $f_{pe}$ measurements; the latter one is computed using the plasma density obtained from the spacecraft potential (see panel (e)). Panel (f) shows the reconstructed magnetic field {\it fff} spectrum, which is very similar to the measured one (panel (b)).

\begin{figure*}
\centering
\includegraphics[width=1\textwidth]{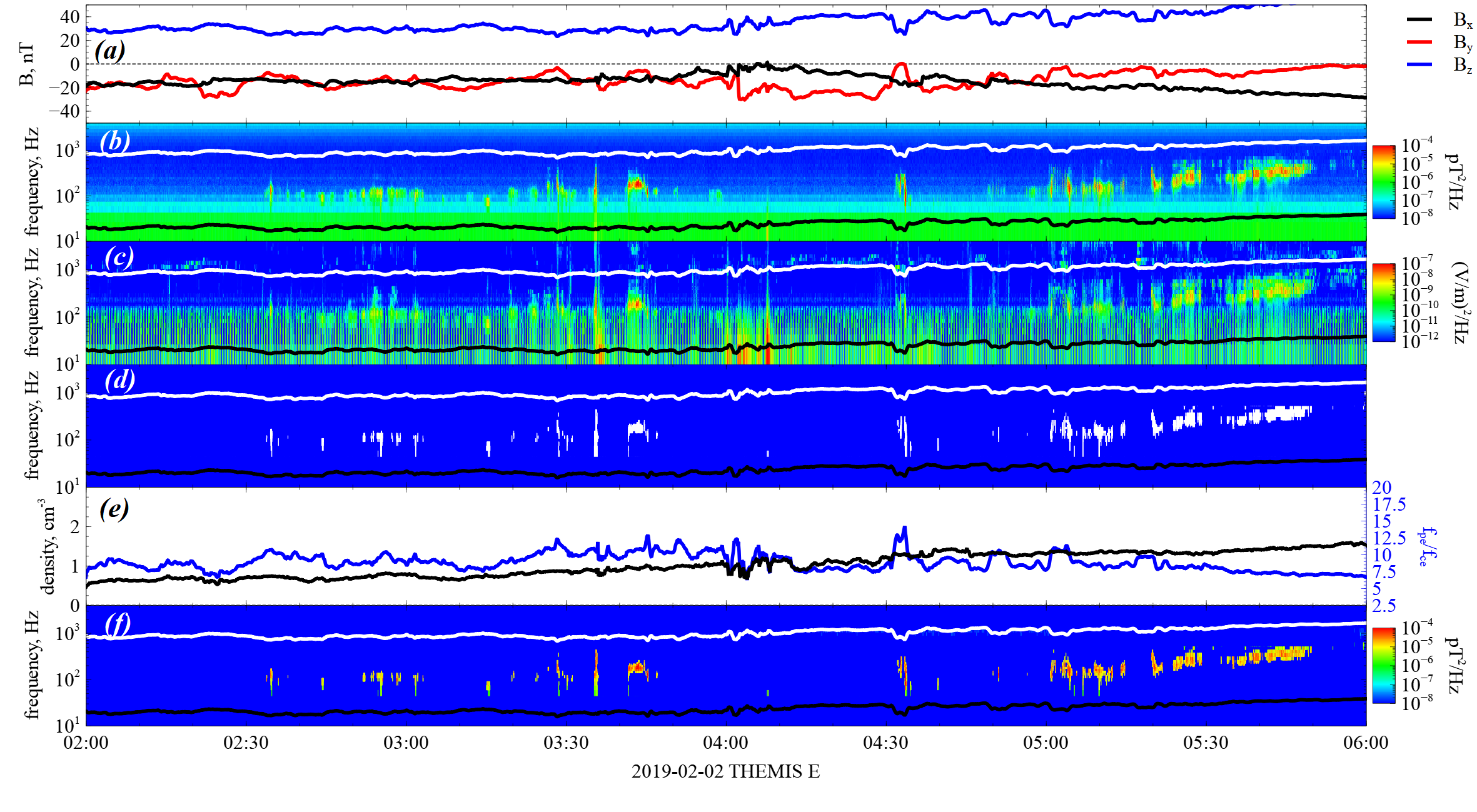}
\caption{Example of a {\it bad} event from THEMIS E measurements with (a) magnetic filed (GSM coordinates), magnetic field spectra density from {\it fff} (b), electric field spectra density from {\it fff} (c), frequency domains of whistler-mode waves determined from magnetic field spectra (d) plasma density and $f_{pe}/f_{ce}$ ratio (e), reconstructed magnetic field spectra density (f). White and blue regions in panel (d) can be considered as having values $1$ (waves) and $0$ (noise); then ``multiplication'' of (d) and (c) panels gives the electric field spectra associated with whistler-mode waves and can be further recalculated into magnetic field spectra shown in panel (f). } \label{fig2}
\end{figure*}

Figure \ref{fig2} shows the application of this reconstruction procedure to THEMIS E measurements with a significantly underestimated magnetic field intensity in the {\it fff} dataset. Panel (a) shows that $B_x$, $B_y$ are comparable with $B_z$, i.e., THEMIS is at some distance from the equator, but still quite close to the nominal wave source region. Panel (b) shows magnetic field spectrum from the {\it fff} dataset with clearly identifiable whistler-mode waves. Using frequency ranges of these whistler-mode wave bursts, we clean the electric field spectrum: compare panels (c) and (d). Then, using local $f_{ce}$ and $f_{pe}$ (see panel (e)), we recalculate magnetic field spectrum from the cleaned electric field spectrum. Panel (f) shows results of such a recalculation: the magnetic field intensity significantly exceeds those from the original {\it fff} dataset (compare with panel (b)).

\section{Statistics of the Search-coil magnetometer measurements}\label{sec:statistics}
The proposed method of electric-to-magnetic field recalculation has several sources of uncertainties that are discussed below. 
To estimate its overall efficiency and accuracy, we would like to perform statistical analysis of the efficiency of the proposed method and determine the correction factor between actual and measured wave intensity. This factor can be used for further statistical electric-to-magnetic field recalculations, without analysis of individual events.
Thus, we pick up $\sim 100$ hours of whistler-mode observations by THEMIS E and by THEMIS D from each year within $2015 - 2022$. 

\begin{figure*}
\centering
\includegraphics[width=1\textwidth]{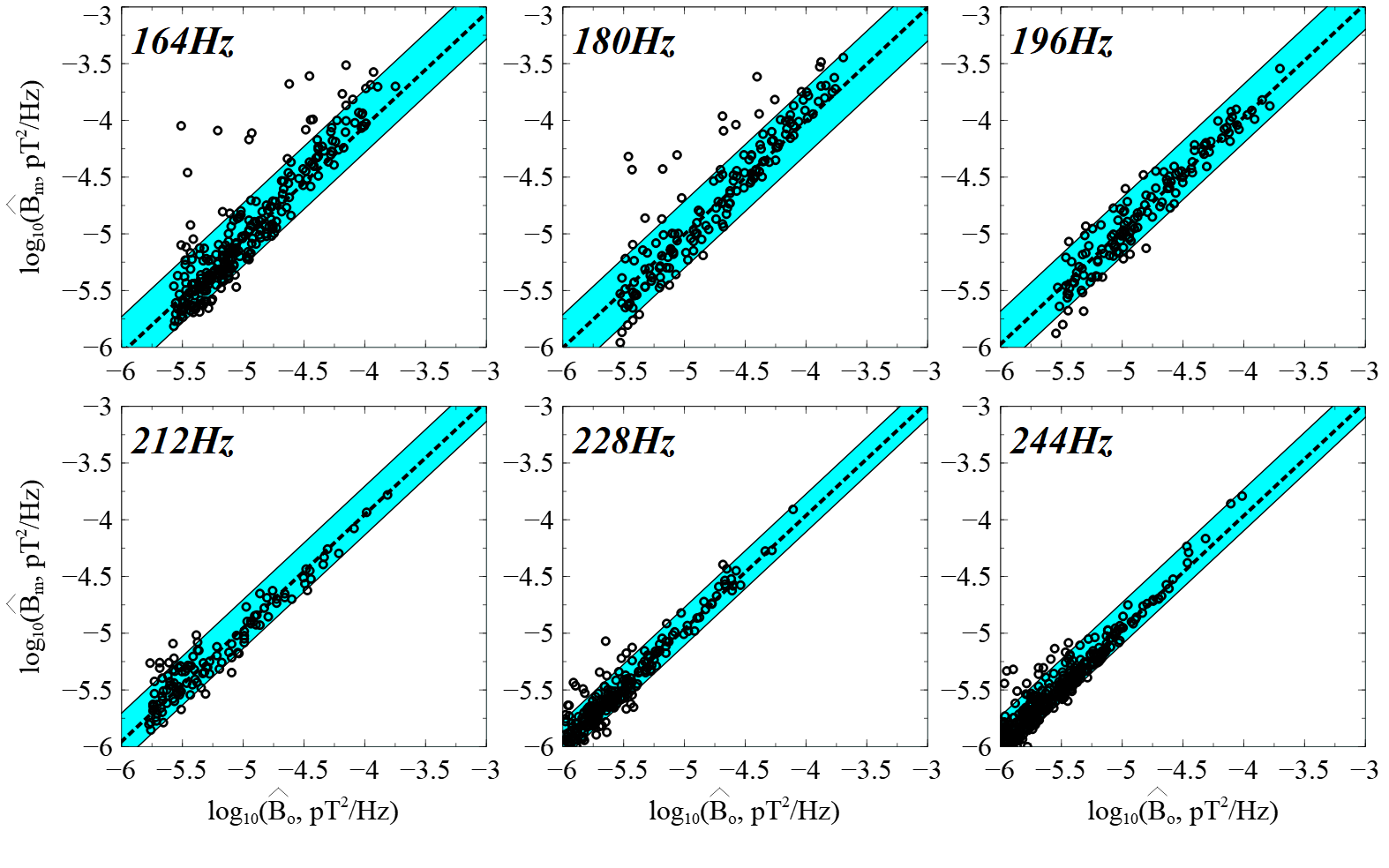}
\includegraphics[width=1\textwidth]{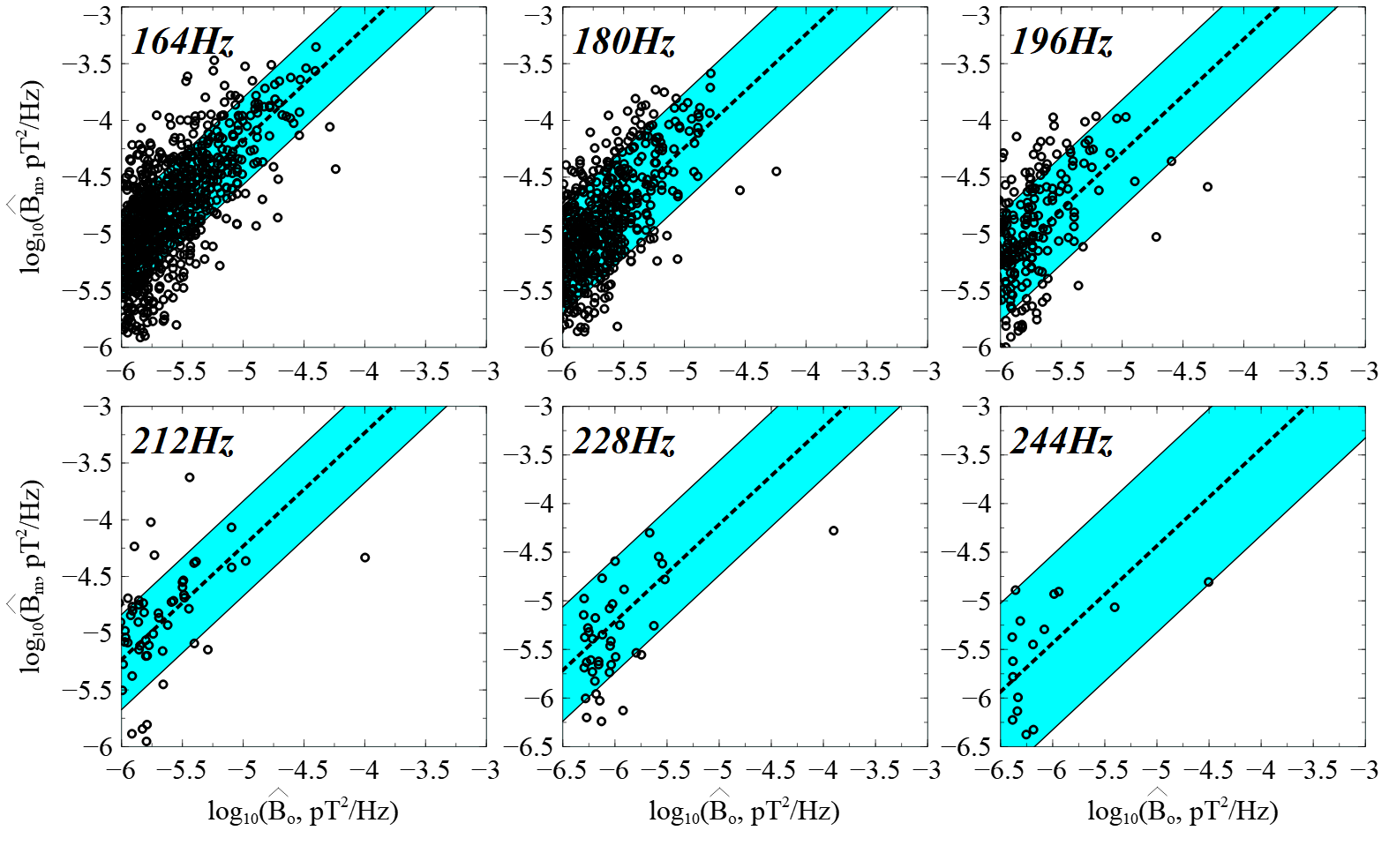}
\caption{Calculation of the parameter $C_{\omega} = \log(B_{\omega,o}^2/B_{\omega,m}^2)$ ({\it good} event from THEMIS A: top panels, {\it bad} event from THEMIS E: bottom panels). \label{fig3a}}
\end{figure*}

We started by considering one-day-long datasets. For each frequency bin in the {\it fff} dataset, we compiled a scatter plot of observed (measured) spectral intensity $B_{\omega,o}^2(f)$ versus a reconstructed model (recalculated) spectral intensity $B_{\omega,m}^2(f)$. Then, we excluded outliers (effectively selecting a `diagonal' strip on the $(B_{\omega,o}^2, B_{\omega,m}^2)$ plane, see Figure \ref{fig3a}). We used the following method to remove the outliers: First, we computed the percentage of the number $N_C$ of measurements for which $\log_{10}(B_{\omega,o}^2) < C_{\omega, try} + \log_{10}(B_{\omega,o}^2)$ as a function of $C_{\omega, try}$. That function increases from $0$ at large negative $C_{\omega, try}$ to $1$ at large positive $C_{\omega, try}$. The boundary of the blue strips in each panel of Figure (\ref{fig3a}) is chosen such that $d N_C/ d C_{\omega, try} = 0.6$. Then, we defined the recalculation parameter as $C_{\omega} = \log(B_{\omega,o}^2/B_{\omega,m}^2)$ for an individual day/individual frequency using the least mean square approach for the points inside the strip. 

The top panels show a significant  correlation of $B_{o}$ and $B_{m}$ for THEMIS A measurements. The difference between these two wave magnitudes are due to the uncertainty in the methodology and the underestimation of the electric field wave intensity in the {\it fff} dataset. The bottom panels show a much weaker correlation of $B_{o}$ and $B_{m}$ for THEMIS E measurements with larger differences between measured and modeled magnetic field magnitudes. Such weaker correlation and larger difference in absolute values of $B_{o}$ and $B_{m}$ are expected because $B_{o}$ for THEMIS E does not match the real wave intensity due to technical problems with the search-coil magnetometer.  

\begin{figure*}
\centering
\includegraphics[width=1\textwidth]{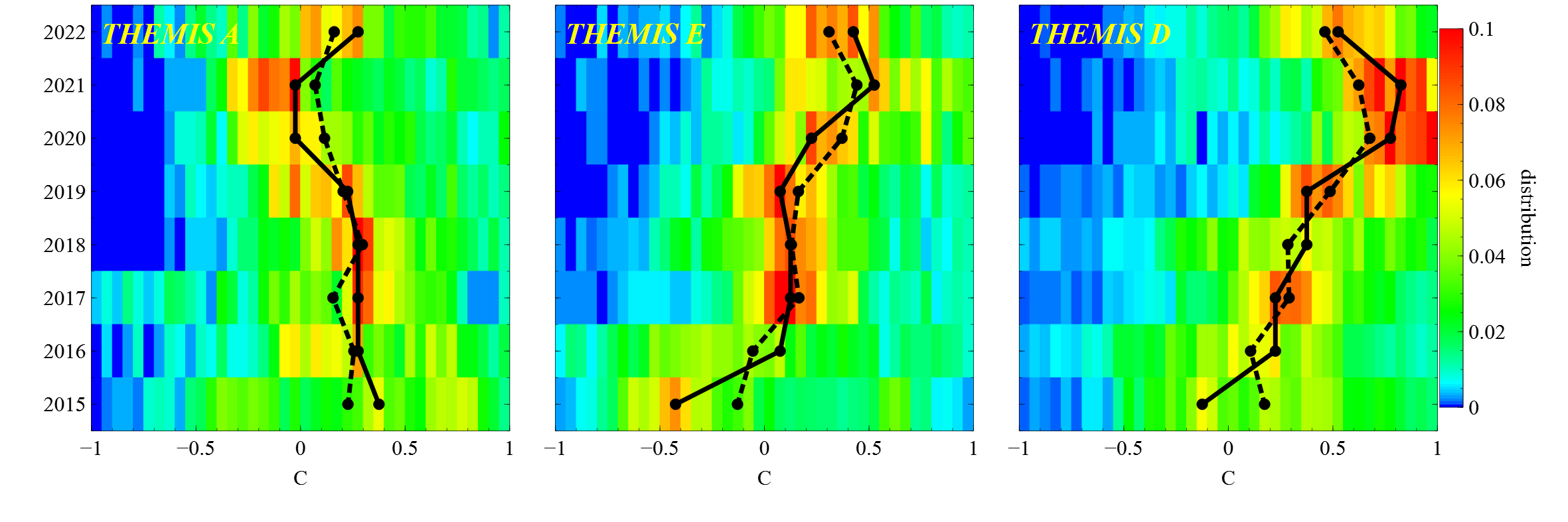} 
\caption{The probability distribution function of $C$ for different years of THEMIS A (left), THEMIS E (center), and THEMIS D (right) search-coil magnetometer measurements. In each year we analyze $\sim 100$ hours of whistler-mode waves ($\sim 1$ month of observations). Altogether, we had $38,125,369$ timestamps distributed over $310$ days in 2015-2022 for THEMIS E, $47,435,522$ timestamps distributed over $509 $ days in 2015-2022 for THEMIS D, and $19,957,851$ timestamps distributed over $246$ days in 2015-2022 for THEMIS A.  Solid and dashed lines show peak value and mean value of $C$-distribution. \label{fig3b}}
\end{figure*}

To quantify the statistics of search-coil magnetometer measurements, we integrated parameters $C_{\omega}$ over the corresponding frequency range (from $f_{ce}/40$ to $f_{ce}$ at that concrete time moment and over the year of measurements) to obtained parameter $C$. The left panel of Figure \ref{fig3b} shows parameter $C$ for THEMIS A data, where $C$ is centered around $\approx0.25$ and remains essentially unchanged throughout 2015-2022. In contrast to results for THEMIS A, distributions of $C$ for THEMIS E and D strongly vary with years: starting with $C\approx -0.25$ and $C\approx 0$ in 2015 and drifting to larger $C$ values at 2022. Note that the absolute value of $C$ can be affected by the particularities by electric and magnetic field measurements, but this value should remain the same for a correctly operating instrument. For example, for THEMIS A, the value of $B_m^2$ recalculated from {\it fff} electric field data should be corrected by the factor $\times 2\approx 10^{0.25}$ to match measured wave magnetic field during normal operation of search-coil magnetometers.

In Table 1, we assembled the mean and peak values of the $C$-distributions for THEMIS D and E for 2015-2022. These values of $C$ for different years should be used for statistical recalculations of {\it fff} magnetic fields for whistler-mode waves, if, for some reason, the {\it fff} electric field data are not available. For example, for THEMIS E such a recalculation assumes the measured $B_o^2$ should be multiplied to $\approx \times 1.5$ factor at 2016 and to $\approx \times 3$ factor in 2021.

 
Let us briefly discuss several uncertainties of the reconstruction technique based on the recalculation of the electric field {\it fff} spectra to the magnetic field {\it fff} spectra. {\bf First}, this technique assumes a  field-aligned wave propagation, since we cannot estimate the propagation angle from the electric field {\it fff} dataset. This assumption works well for the most intense near-equatorial whistler-mode waves. However, this technique may fail for oblique waves \cite{Agapitov13:jgr,Li16:statistics} observed around the equatorial source region \cite{Li16} or at middle and high latitudes \cite{Agapitov18:jgr}. The wave obliquity significantly changes the electric-to-magnetic field ratio for angles above the Gendrin angle, and is especially important for waves propagating around the resonant cone angle \cite<see the review>{Artemyev16:ssr}.

{\bf Second}, this technique utilizes the cold plasma dispersion relation for recalculation of the magnetic wave field from the electric wave fields. The cold plasma dispersion should work well for quasi-parallel propagating waves, but does not provide an adequate description of wave dispersion for very oblique waves \cite<e.g.,>{Ma17}.

\begin{table}
\centering
\begin{tabular}{ l||c|c|c|c| } 
year & mean ThE & peak ThE & mean ThD & peak ThD \\
 \hline
 2015 & $-0.13\pm0.46$ & $-0.425$ & $-0.17\pm0.41$ & $-0.125$  \\ 
 2016 & $-0.06\pm0.46$ & $0.075$ & $0.11\pm0.40$ & $0.225$  \\ 
 2017 & $0.16\pm0.3$ & $0.125$ & $0.29\pm0.32$ & $0.225$  \\ 
 2018 & $0.13\pm0.34$ & $0.125$ & $0.28\pm0.4$ & $0.375$  \\  
 2019 & $0.16\pm0.31$ & $0.075$ & $0.49\pm0.3$ & $0.375$  \\   
 2020 & $0.37\pm0.32$ & $0.225$ & $0.68\pm0.3$ & $0.775$  \\   
 2021 & $0.44\pm0.33$ & $0.525$ & $0.62\pm0.31$ & $0.825$  \\    
 2022 & $0.31\pm0.3$ & $0.425$ & $0.46\pm0.31$ & $0.525$  \\    
 \hline
\end{tabular}
 \caption{Mean and peak values of $C$-distributions for THEMIS E and D from Fig. \ref{fig3b}}
\label{table}
\end{table}

{\bf Third}, we use the spacecraft potential to estimate the plasma density for the wave dispersion relation \cite{Nishimura13:density}, and this method may have significant uncertainties for systems with rarefied hot plasma \cite<see discussion in>[and references therein]{Andriopoulou16,Andriopoulou18}.

{\bf Fourth}, we use the electric field {\it fff} spectra that combine contributions of whistler-mode waves, various electrostatic waves, and electric field noise. Therefore, the actual whistler-mode wave electric field can be overestimated. On the other hand, and somewhat balancing this overestimation, {\it fff} includes only two of the three electric field components (in the spacecraft frame) and thus can underestimate whistler-mode wave electric field.

\section{Conclusions}\label{sec:conclusions}
In this technical report, we proposed and verified a technique for the reconstruction of whistler-mode magnetic field intensity for the THEMIS E, D {\it fff} datasets. This technique is based on using the electric field {\it fff} dataset and the cold-plasma dispersion relation. The key element of the technique consists in distinguishing (frequency, time) domains of whistler-mode waves using magnetic field {\it fff} spectra with underestimated wave intensity. This element allows us to separate whistler-mode waves in the electric field {\it fff} spectra and minimize the uncertainty associated with various electrostatic modes observed within the whistler-mode wave frequency range. 

Using THEMIS A with a well-operating search-coil magnetometer, we show the general accuracy of the proposed reconstruction technique, whereas  statistics of the THEMIS E, D {\it fff} datasets provide an averaged recalculation factor $C$ mostly useful for statistical reconstruction of magnetic field spectra after 2016.

\section*{Appendix: Z-score Method}
\begin{figure*}
\centering
\includegraphics[width=1\textwidth]{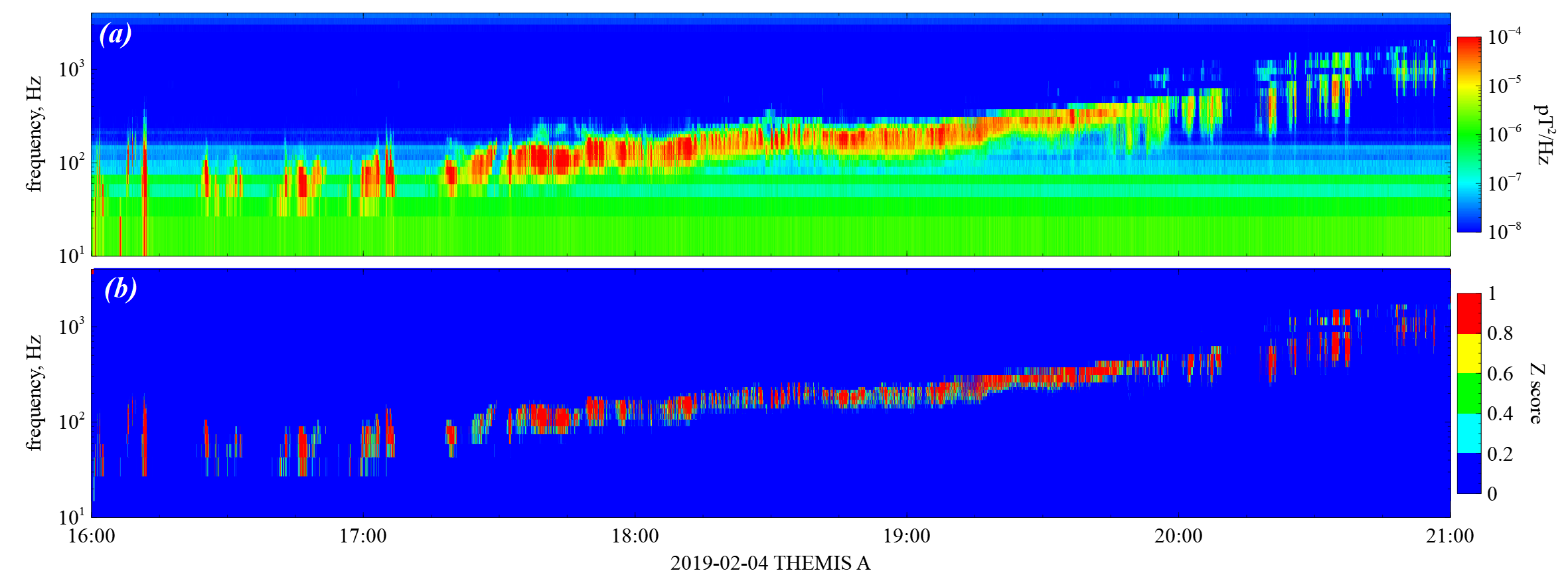} 
\caption{Illustration of $Z$-method used to identify the waves. \label{fig6_new}}
\end{figure*}

To reconstruct the magnetic field data, we needed a way to isolate the high-intensity whistler waves from the low intensity noise. The low-intensity noise constitutes a vast majority of the data, and the whistler waves are fairly sparse (essentially, they are outliers). Thus, we used the Z-score method to find outliers. For each point x, we calculate its Z-score as $ Z = \left(x-\mu \right) /\sigma$, where $\mu$ is the mean and $\sigma$ is the standard deviation. Typically, the mean and standard deviation would be taken from the whole dataset. However, for the purposes of this study, we are calculating mean and standard deviation using the data from the given channel, and the channels directly adjacent to the chosen chanel when calculating Z-scores. We found that the whistler waves start to become apparent around a score of +0.2, as seen in Fig.~\ref{fig6_new}. We chose $Z=+0.8$ for the threshold, but, as can be seen from Fig.~\ref{fig6_new}, the whistlers' identification is not too sensitive to the exact threshold value. After determining the desired data points using the Z-score method, some additional cleaning measures were applied. Occasionally, a frequency channel containingundesired values will yield a sizable number of false positives. In a proper isolation, a channel typically only has less than 10\% of its data with $Z > +0.8$. So, if a channel has more than 10\% of points with $Z > +0.8$, we remove the entire channel from consideration. Then, the final cleaning measure removes all leftover points that are outside of the range from \(f_{lh}\) to \(f_{pe}\).

\acknowledgments
We acknowledge support by NASA awards 80NSSC20K1578, 80NSSC23K1038, and 80NSSC25K7749 (A.V.A., D.L.V.). We acknowledge the support of NASA contract NAS5-02099 for use of data from the THEMIS Mission (A.V.A., D.L.V., V.A.).

\subsection*{Open Research}\noindent
THEMIS data is available at \url{http://themis.ssl.berkeley.edu}. THEMIS data access and processing was done using SPEDAS V4.1, see \citeA{Angelopoulos19}. The codes used for the present report with corresponding documentation are available at \url{https://github.com/DeclanDoesDev/THEMIS-SCM-Reconstructor}



\end{document}